\documentclass[showpacs,preprintnumbers,amsmath,amssymb,floatfix]{revtex4}
\usepackage{graphicx}
\usepackage{dcolumn}
\usepackage{bm}
\usepackage{epsfig}
\usepackage{verbatim}
\usepackage{marvosym}
\usepackage{supercite}

\newcommand{\hs}{\hspace*{1mm}}
\newcommand{\beq}{\begin{eqnarray}}
\newcommand{\eeq}{\end{eqnarray}}
\newcommand{\no}{\nonumber}
\newcommand{\bs}{\boldsymbol}

\begin{document}
\setlength{\unitlength}{1mm}
\title{Density and Stability in Ultracold Dilute Boson-Fermion Mixtures}

\author{Steffen R{\"o}thel}
\email{steffen.roethel@uni-muenster.de}
\affiliation{Westf{\"a}lische Wilhelms-Universit{\"a}t M{\"u}nster,
Wilhelm-Klemm-Stra{\ss}e 9, 48149 M{\"u}nster, Germany}
\author{Axel Pelster}
\email{axel.pelster@uni-duisburg-essen.de}
\affiliation{Universit{\"a}t Duisburg-Essen, Campus Duisburg, Fachbereich Physik, 
Lotharstra{\ss}e 1, 47048 Duisburg, Germany}     

\date{\today}

\begin{abstract}
We analyze in detail recent experiments on ultracold dilute $^{87}$Rb--$^{40}$K mixtures in Hamburg and in Florence within
a mean-field theory. To this end we determine how the stationary bosonic and fermionic density profiles in this mixture
depend in the Thomas-Fermi limit on the respective particle numbers. Furthermore, we investigate 
how the observed stability of the Bose-Fermi mixture with respect to collapse is 
crucially related to the value of the interspecies s-wave scattering length.
\end{abstract}
\maketitle
\section{Introduction}
Six years after the first experimental achievement of Bose-Einstein condensation 
(BEC) of trapped atomic gases in 1995 fermionic atomic 
gases were brought together with bosonic atoms to quantum degeneracy in a $^7$Li--$^6$Li mixture 
\cite{Truscott,Schreck}, $^{23}$Na--$^6$Li mixture \cite{Gorlitz}, and $^{87}$Rb--$^{40}$K mixture 
\cite{Inguscio}. In contrast to a pure Bose system, quantum degeneracy in a Fermi system with only one 
spin component means for $T\ll T_F = E_F/k_B$
that all energy states below the Fermi energy $E_F$ are occupied with one fermion 
each, whereas all states above $E_F$ remain empty. The main 
problem to achieve quantum degeneracy in a Fermi gas is the inability of fermions to be directly 
evaporatively cooled. Due to the Pauli exclusion principle, fermions in the same spin 
polarized hyperfine state  are not allowed to be close together, so that they can not collide via short-range 
contact interaction to rethermalize the gas during the evaporative cooling. This handicap was 
circumvented in the experiment of Ref.~\cite{DeMarco}, where a mixture with two different spin states 
of $^{40}$K was simultaneously evaporated by mutual cooling. In combination with a Bose gas the fermions 
are sympathetically cooled by elastic interactions with the bosons in the overlapping region \cite{Inguscio,Sengstock}.

Beside the exploration of quantum degeneracy, one is also interested in studying how the two-particle 
interaction influences the system properties. Mixtures with a strong interspecies interaction, with 
the prominent example of $^4$He--$^3$He liquid \cite{Yokoshi,Capuzzi}, lead to new phenomena like phase 
separation or BEC-induced interactions between fermions. Depending on the nature of the interspecies 
interaction, a repulsion between bosons and fermions tends to a demixing in order to minimize the overlapping 
region \cite{Bongs}, 
whereas in the case of an attraction the mixture can collapse as long as the particle numbers are 
sufficiently large \cite{Sengstock,Inguscio-Sc}. The possibility of superfluidity in a Fermi gas, 
especially the predicted BEC-BCS crossover between BCS-type superfluidity of Cooper pairs of fermionic atoms 
and BEC of molecules was recently probed \cite{Regal,Zwierlein}. A Feshbach resonance was 
used to tune the interaction strength between fermionic atoms of two different spin states, characterized 
by the s-wave scattering length $a$, from effectively repulsive $(a>0)$ to attractive $(a<0)$. On the BEC 
side ($a>0$) of the magnetic field resonance, there exists a weakly bound molecular state. Fermionic atoms are bound 
into bosonic molecules which can condense at sufficient low temperatures. On the BCS side ($a<0$) of the 
resonance two fermions with different spin states form a loosely bounded Cooper pair. The condensation of 
fermionic atom pairs was observed on both the BEC and the BCS side of the Feshbach resonance in the 
experiment \cite{Regal,Zwierlein}. Furthermore, the system properties were observed to vary smoothly in the BEC-BCS 
crossover regime. An alternative and complementary access to Fermi superfluidity is expected from quantum 
degenerate Bose-Fermi mixtures where an effective interaction between fermions is mediated by the bosons 
\cite{Viverit,Stoof}, similarly to phonons in a solid-state superconductor.

Another recent and fast growing field is the investigation of ultracold boson-fermion mixtures trapped 
in an  optical lattice which is created by standing waves of the electric field of counterpropagating 
laser beams. The atoms can be confined to different lattice sites and, by varying the laser intensity, 
their tunneling to neighboring sites as well as the strength of their on-site repulsive interactions 
can be controlled. In the case of a pure ultracold Bose-Einstein condensate with repulsive interaction, 
held in a three-dimensional optical lattice potential, a quantum phase transition from a superfluid to a 
Mott insulator phase was observed as the depth of the lattice is increased leading to a suppression of 
the tunneling between neighboring lattice sites \cite{M-Greiner}. The presence of fermionic atoms together 
with the Bose-Einstein condensate makes the system more complex and richer in its behavior at low 
temperatures. It has been predicted that novel quantum phases in the strong-coupling regime occur which 
involve the pairing of fermions with one or more bosons or bosonic holes, respectively, when the 
boson-fermion interaction is attractive or repulsive \cite{Lewenstein}. Depending on the physical parameters 
of the system, these composite fermions may appear as a normal Fermi liquid, a density wave, a superfluid, 
or an insulator with fermionic domains. Instead of varying the lattice potential depth the transition from 
a superfluid to a Mott insulator in bosonic $^{87}$Rb can be shifted towards larger lattice depth by 
adding fermionic $^{40}$K  which interacts attractively with rubidium and therefore increases the 
effective lattice depth \cite{Ospelkaus,Guenter}.

Ultracold trapped boson-fermion mixtures were investigated with respect to a demixing of the components \cite{Nygaard,Feldmeier,Roth} and to a collapse due to the interspecies attraction \cite{Miyakawa,Feldmeier,Roth,Chui,Ryzhov}. Furthermore, the time-dependent dynamics of the collapse \cite{Adhikari} and finite temperature effects on the stability in a boson-fermion mixture were also studied \cite{Liu}. Our theoretical investigation is based on the experiments with a $^{87}$Rb--$^{40}$K boson-fermion mixture 
in a harmonic trap, which were performed in Hamburg \cite{Sengstock} and in Florence \cite{Inguscio,Inguscio-Sc}. 
\begin{table}[t]
\centering
\begin{minipage}[t]{.93\textwidth}
\centering
\begin{tabular}[t]{p{45mm}ll} \hline\hline
 & \hspace*{-6mm} Hamburg Experiment \hspace*{3mm} & Florence Experiment \\\hline
mass of $^{87}$Rb atom & \multicolumn{2}{c}{$m_B = 14.43\cdot 10^{-26}$ kg} \\
mass of $^{40}$K atom & \multicolumn{2}{c}{$m_F = 6.636\cdot 10^{-26}$ kg}  \\
\mbox{s-wave scattering length} (bosons $\leftrightarrow$ bosons) & \multicolumn{2}{c}{$a_{BB} = (5.238\pm 0.002)$ nm} \\
\mbox{s-wave scattering length} (bosons $\leftrightarrow$ fermions) & 
$a_{BF} = -15.0$ nm & $a_{BF} = (-20.9\pm 0.8)$ nm \\
radial trap frequency (bosons) \hspace*{10mm}& $\omega_{B,r} = 2\pi\cdot 257$ Hz & $\omega_{B,r} = 2\pi\cdot 215$ Hz \\
axial trap frequency (bosons) & $\omega_{B,z} = 2\pi\cdot 11.3$ Hz & $\omega_{B,z} = 2\pi\cdot 16.3$ Hz \\
radial trap frequency (fermions) & $\omega_{F,r} = 2\pi\cdot 379$ Hz & $\omega_{F,r} = 2\pi\cdot 317$ Hz \\
axial trap frequency (fermions) & $\omega_{F,z} = 2\pi\cdot 16.7$ Hz & $\omega_{F,z} = 2\pi\cdot 24.0$ Hz \\
number of bosons & $N_B = 10^6$ & $N_B = 2\cdot 10^5$ \\
number of fermions  & $N_F = 7.5\cdot 10^5$ & $N_F = 3\cdot 10^4$ \\\hline\hline
\end{tabular}
\caption{List of parameters  of the experiments with a $^{87}$Rb--$^{40}$K boson-fermion mixture. 
The values are taken from the experiments in Hamburg \cite{Sengstock} and in Florence 
\cite{Inguscio,Inguscio-Sc,Ferlaino}.}
\label{tab:exp-parameter}
\end{minipage}
\end{table}
After cooling down at a temperature below 1~$\mu$K, a condensate of $10^6$ $(2\cdot 10^5)$ $^{87}$Rb atoms coexisting 
with $7.5 \cdot 10^5$ $(3\cdot 10^4)$ $^{40}$K atoms with a quantum degeneracy of about $T/T_F=0.1$ $(T/T_F=0.3)$ was 
achieved in the Hamburg (Florence) experiment. The parameters of both experiments are summarized in Table 
\ref{tab:exp-parameter}. The distinct values for the interspecies s-wave scattering length $a_{BF}$ employed to describe each experiment are worth a detailed explanation since this parameter is of great importance for the system, 
especially for the stability of the mixture against collapsing. An overview of different values for $a_{BF}$ 
and their determination method along with a reference are shown in Table \ref{tab:a-B-F}.
\begin{table}[t]
\centering
\begin{minipage}[t]{.93\textwidth}
\centering
\begin{tabular}[t]{l m{87mm}l} \hline\hline
$a_{BF}/a_\mathrm{Bohr}$ & Method of determination & Reference (year) \\\hline
$-261_{-159}^{+170}$ & measurement of  the elastic cross section for collisions between $^{41}$K 
and $^{87}$Rb in different temperature regimes and following mass scaling to the fermionic $^{40}$K 
isotope & \cite{Ferrari} (2002)\\
$-330_{-100}^{+160}$ & measurement of the rethermalization time in the mixture in \cite{Inguscio,Inguscio-Sc} 
after a selectively heating of $^{87}$Rb & \cite{Inguscio} (2002)\\
$-410_{-91}^{+81}$ & measurement of the damping of the relative oscillations of $^{40}$K and $^{87}$Rb in a 
magnetic trap & \cite{Inguscio-Sc} (2002)\\
$-395\pm 15$ & mean-field analysis of the stability of the mixture in \cite{Inguscio,Inguscio-Sc} & 
\cite{Ferlaino} (2003)\\
$-281\pm 15$ &  magnetic Feshbach spectroscopy of an ultracold mixture of $^{40}$K and $^{87}$Rb atoms & 
\cite{Inouye} (2004)\\
$250\pm 30$ & cross dimensional thermal relaxation in a mixture of $^{40}$K and $^{87}$Rb atoms after a 
increase of the radial confinement of the magnetic trap, here only $|a_{BF}/a_0|$ & \cite{Goldwin} (2004)\\
$-284$ & mean-field analysis of the stability, based on \cite{Chui}, of the mixture in \cite{Sengstock} & 
\cite{Sengstock} (2006)\\
$-205\pm 5$ & extensive magnetic Feshbach spectroscopy of an ultracold mixture of $^{40}$K and $^{87}$Rb atoms & 
\cite{Errico} (2006)
\\\hline\hline
\end{tabular}
\caption{List of several published values of the s-wave scattering length between $^{87}$Rb and $^{40}$K 
including their determination method and their reference.}
\label{tab:a-B-F}
\end{minipage}
\end{table}
A comparison of the incompatible values for $a_{BF}$ shows the need of further investigation in this field.

Now we give an outline of the contents of the paper. At first,  we develop in Section \ref{deriv-GP} a 
mean-field theory for an ultracold dilute gaseous boson-fermion mixture within the functional integral 
approach to many-body theory. By splitting the Bose fields into background fields and fluctuation fields, we 
derive an effective action of the Bose subsystem without integrating out the fermionic degrees of freedom. 
Its extremization at zero temperature yields two coupled equations of motion, one for the condensate wave 
function and another one for the Green function of the fermions. By evaluating the fermionic Green function 
for a stationary BEC within the semiclassical approximation, we obtain how the time-independent Gross-Pitaevskii 
equation for the condensate wave function is modified in such a mixture.

In Section \ref{sol-GP} we apply the Thomas-Fermi approximation by neglecting the kinetic energy of the bosons 
and obtain an algebraic Gross-Pitaevskii equation. With the help of its solution we determine the density 
profiles of both components in a $^{87}$Rb--$^{40}$K mixture, where the contact interaction is repulsive between 
the bosons and attractive between both components.

Finally,  we investigate in Section \ref{Stability-collaps} the stability of the Bose-Fermi mixture with respect 
to collapse by numerically evaluating  the effective action for a trial Gaussian density profile of 
the condensate. We compare our results, 
which strongly depend on the value of the Bose-Fermi s-wave scattering length, with the experiments on 
$^{87}$Rb--$^{40}$K 
mixtures in Hamburg and Florence. The good agreement of our theoretical results with the measurement in the Florence experiment allows us to fit $a_{BF}$ which describes the Hamburg data quite well, but remains incompatible with the value used in Florence.
\section{Derivation of Gross-Pitaevskii Equation}\label{deriv-GP}
In this section we summarize briefly the mean-field theory of boson-fermion mixtures \cite{Chui,Stoof} 
and derive within the functional integral formalism  a coupled set of differential equations for the 
condensate wave function and the fermionic Green function. 
\subsection{Grand-Canonical Partition Function}
We consider a dilute gaseous mixture of ultracold  bosonic and fermionic atoms. In order to obtain quantum 
statistical quantities for such a Bose-Fermi mixture, we use  the grand-canonical partition function in 
the functional integral formalism. Thus, we integrate over all possible Bose fields $\psi_B^*(\mathbf{x},\tau)$, 
$\psi_B(\mathbf{x},\tau)$ and Fermi fields $\psi_F^*(\mathbf{x},\tau)$, $\psi_F(\mathbf{x},\tau)$, which are 
weighted by a Boltzmann factor \cite{Stoof,Greiner,Kleinert-Pfad}:
\beq\label{Z}
\mathcal{Z}=\oint\mathcal{D}\psi_B^*\oint\mathcal{D}\psi_B\oint\mathcal{D}\psi_F^*\oint\mathcal{D}\psi_F
\hs e^{-\mathcal{A}[\psi_B^*,\psi_B,\psi_F^*,\psi_F]/\hbar}.
\eeq
The complex fields $\psi_B^*(\mathbf{x},\tau)$, $\psi_B(\mathbf{x},\tau)$ represent the bosons and are periodic on the 
imaginary time interval \([0,\hbar\beta]\), whereas the fermions are described by Grassmann fields 
$\psi_F^*(\mathbf{x},\tau)$, $\psi_F(\mathbf{x},\tau)$ which are antiperiodic on this interval:
\beq\label{periodic}
\psi^*(\mathbf{x},\hbar\beta)=\epsilon\,\psi^*(\mathbf{x},0), 
\qquad
\psi(\mathbf{x},\hbar\beta)=\epsilon\,\psi(\mathbf{x},0).
\eeq
Here \(\epsilon=\pm1\) holds for bosons and fermions, respectively. The total euclidean action of a Bose-Fermi 
mixture consists of three parts:
\beq\label{A-eucl}
\mathcal{A}[\psi_B^*,\psi_B,\psi_F^*,\psi_F]=\mathcal{A}_B[\psi_B^*,\psi_B]+\mathcal{A}_F[\psi_F^*,\psi_F]+
\mathcal{A}_{BF}[\psi_B^*,\psi_B,\psi_F^*,\psi_F].
\eeq
The first term describes the bosonic component of the mixture:
\beq\label{A-B}
\mathcal{A}_B[\psi_B^*,\psi_B] &=& \int_0^{\hbar\beta}d\tau \int d^3x\,\psi_B^*(\mathbf{x},\tau)
\left[\hbar\frac{\partial}{\partial\tau}-\frac{\hbar^2}{2m_B}\Delta
+V_B({\mathbf{x}})-\mu_B +\frac{g_{BB}}{2}\,|\psi_B(\mathbf{x},\tau)|^2\right] \psi_B(\mathbf{x},\tau).
\eeq
It contains the Legendre transform, the kinetic energy, the external trap potential \(V_B(\mathbf{x})\), 
the chemical potential $\mu_B$ to fix the boson number, and the strength $g_{BB}= 4 \pi \hbar^2 a_{BB} / m_B $
of the contact interaction   between two bosons with the s-wave scattering length $a_{BB}$.
As we deal with a dilute gas, an interaction between more than two particles is negligible, and we can restrict
ourselves to the short-range contact interaction. Since the Pauli principle forbids fermions in the same 
hyperfine state to be close together and therefore to collide via contact interaction, we can write the 
corresponding action term for the fermionic component of the mixture as
\beq\label{A-F}
\mathcal{A}_F[\psi_F^*,\psi_F] = \int_0^{\hbar\beta}d\tau \int d^3x\,\psi_F^*(\mathbf{x},\tau)
\left[\hbar\frac{\partial}{\partial\tau}-\frac{\hbar^2}{2m_F}\Delta
+V_F(\mathbf{x})-\mu_F\right] \psi_F(\mathbf{x},\tau),
\eeq
where \(V_F(\mathbf{x})\) and $\mu_F$ denote the external trap potential and the chemical potential for fermions. 
As we assume a situation, where the bosonic and fermionic atoms cannot be transformed into each other, each of 
both species has its own chemical potential. The last term of the euclidean action (\ref{A-eucl})
\beq\label{A-BF}
\mathcal{A}_{BF}[\psi_B^*,\psi_B,\psi_F^*,\psi_F] = g_{BF} \int_0^{\hbar\beta}d\tau \int d^3x 
\,|\psi_B(\mathbf{x},\tau)|^2\,|\psi_F(\mathbf{x},\tau)|^2
\eeq
describes the contact interaction between bosons and fermions, where its strength $g_{BF}$ is related to 
the s-wave scattering lengths $a_{BF}$ via \cite{Pethick,Pitaevskii}
\beq\label{g}
g_{BF} = 2\pi\hbar^2 a_{BF}\, \frac{m_B+m_F}{m_B\,m_F}.
\eeq
\subsection{Background Method}
In order to account for the fact that the bosons in the mixture can condense, we apply the background method of 
field theory \cite{De Witt,Jackiw,Kleinert-Pfad} and split the bosonic fields 
$\psi_B^*(\mathbf{x},\tau)$, $\psi_B(\mathbf{x},\tau)$ into two parts:
\beq\label{cond+thermal-bos}
 \psi_B^*(\mathbf{x},\tau)=\Psi_B^*(\mathbf{x},\tau)+\delta\psi_B^*(\mathbf{x},\tau),
\qquad \psi_B(\mathbf{x},\tau)=\Psi_B(\mathbf{x},\tau)+\delta\psi_B(\mathbf{x},\tau).
\eeq
The first part represents the background fields $\Psi_B^*(\mathbf{x},\tau)$, $\Psi_B(\mathbf{x},\tau)$ 
whose absolute square 
is identified with the density of the condensed bosons. The second part consists of the fluctuation fields 
$\delta\psi_B^*(\mathbf{x},\tau)$, $\delta\psi_B(\mathbf{x},\tau)$ of the Bose gas describing the excited bosons, 
which are not in the 
ground state. Using the decomposition (\ref{cond+thermal-bos}), we expand the euclidean action (\ref{A-eucl}) 
in a functional Taylor series with respect to the Bose fields $\psi_B^*(\mathbf{x},\tau)$, $\psi_B(\mathbf{x},\tau)$  
around 
the background fields $\Psi_B^*(\mathbf{x},\tau)$, $\Psi_B(\mathbf{x},\tau)$. In this work we restrict ourselves to the 
Gross-Pitaevskii theory, i.e.~we consider the euclidean action \(\mathcal{A}\) only up to the zeroth order in the 
fluctuation fields $\delta\psi_B^*(\mathbf{x},\tau)$, $\delta\psi_B(\mathbf{x},\tau)$ which is equivalent to evaluate the 
euclidean action (\ref{A-eucl}) at the background fields:
$\mathcal{A}[\Psi_B^*+\delta\psi_B^*,\Psi_B+\delta\psi_B,\psi_F^*,\psi_F] \approx \mathcal{A}[\Psi_B^*,
\Psi_B,\psi_F^*,\psi_F]$. Thus, the bosonic functional 
integration in Eq.~(\ref{Z}), whose integration measure transforms according $\mathcal{D}\psi_B^*(\mathbf{x},\tau)=
\mathcal{D}\delta\psi_B^*(\mathbf{x},\tau)$ and $\mathcal{D}\psi_B(\mathbf{x},\tau)
=\mathcal{D}\delta\psi_B(\mathbf{x},\tau)$, can be 
dropped. In this way we obtain for the effective action $\Gamma[\Psi_B^*,\Psi_B]
=- \ln\mathcal{Z}[\Psi_B^*,\Psi_B]/\beta$ the result
\beq\label{F8}
\Gamma[\Psi_B^*,\Psi_B] = \frac{1}{\hbar\beta}\,\mathcal{A}_B[\Psi_B^*,\Psi_B]-\frac{1}{\beta} 
\ln \mathcal{Z}_F[\Psi_B^*,\Psi_B],
\eeq
where
\beq\label{Z-F}
\mathcal{Z}_F[\Psi_B^*,\Psi_B] = \oint\mathcal{D}\psi_F^*\oint\mathcal{D}\psi_F\, 
e^{-\mathcal{A}_{F,\mathrm{eff}}[\Psi_B^*,\Psi_B,\psi_F^*,\psi_F]/\hbar}
\eeq
represents the functional integral over the Fermi fields resulting in a pure functional of the Bose background 
fields $\Psi_B^*(\mathbf{x},\tau)$, $\Psi_B(\mathbf{x},\tau)$. The effective euclidean action 
$\mathcal{A}_{F,\mathrm{eff}}[\Psi_B^*,\Psi_B,\psi_F^*,\psi_F]$ depending on the Fermi fields 
$\psi_F^*(\mathbf{x},\tau)$, 
$\psi_F(\mathbf{x},\tau)$ is summarized by
\beq\label{A-F+A-BF}
\mathcal{A}_{F,\mathrm{eff}}[\Psi_B^*,\Psi_B,\psi_F^*,\psi_F] = \int_0^{\hbar\beta}d\tau 
\int d^3x\,\psi_F^*(\mathbf{x},\tau)\left[\hbar\frac{\partial}{\partial\tau}
+\hat H_{F,\mathrm{eff}}(\mathbf{x},\tau)-\mu_F\right] \psi_F(\mathbf{x},\tau),
\eeq
where the effective one-particle Hamilton operator for fermions reads:
\beq\label{Hamilt-F}
\hat H_{F,\mathrm{eff}}(\mathbf{x},\tau) = -\frac{\hbar^2}{2m_F}\Delta+V_F(\mathbf{x})+g_{BF}\,|
\Psi_B(\mathbf{x},\tau)|^2.
\eeq
\subsection{Coupled Equations of Motion}\label{motion-eqs}
The grand-canonical free energy $\mathcal{F}$ is the extremum of the effective action $\Gamma[\Psi_B^*,\Psi_B] $ with 
respect to the background fields $\Psi^*(\mathbf{x},\tau)$, $\Psi(\mathbf{x},\tau)$. Using Eqs.~(\ref{A-B}) and 
(\ref{F8})--(\ref{Hamilt-F}), this yields the time-dependent Gross-Pitaevskii equation:
\beq
\label{B-mot-eq-imag}
\left[\hbar\frac{\partial}{\partial\tau}-\frac{\hbar^2}{2m_B}\Delta+V_B({\mathbf{x}})-\mu_B+g_{BB}\,|
\Psi_B(\mathbf{x},\tau)|^2 
- g_{BF} G_F(\mathbf{x},\tau;\mathbf{x},\tau) \right]\Psi_B(\mathbf{x},\tau) = 0,
\eeq
where the fermionic \mbox{two-point} function is defined as follows:
\beq
G_F(\mathbf{x},\tau;\mathbf{x}',\tau') \equiv \frac{1}{\mathcal{Z}_F[\Psi_B^*,\Psi_B]} \oint\mathcal{D}\psi^*_F 
\oint\mathcal{D}\psi_F\,
\psi_F(\mathbf{x},\tau)\,\psi_F^*(\mathbf{x}',\tau')\, e^{-\mathcal{A}_{F,\mathrm{eff}}[\Psi_B^*,\Psi_B,
\psi_F^*,\psi_F]/\hbar}.
\eeq
This imaginary time-dependent nonlinear Schr{\"o}dinger equation (\ref{B-mot-eq-imag}) of the condensate wave function 
$\Psi_B(\mathbf{x},\tau)$ represents a partial differential equation, where the nonlinear terms are due to 
both interactions. The last term results from the boson-fermion interaction, whereas the other terms have the 
conventional Gross-Pitaevskii form for a condensate \cite{Pitaevskii-orig,Gross-orig}. The fermionic two-point 
function is at the same time a Green function for the fermion fields, thus, it obeys the linear inhomogeneous 
Schr{\"o}dinger equation for fermions
\beq\label{F-mot-eq-imag}
\left[\hbar\frac{\partial}{\partial\tau}+\hat H_{F,\mathrm{eff}}(\mathbf{x},\tau)-\mu_F\right]G_F(\mathbf{x},\tau;\mathbf{x}',\tau') 
= \hbar\,\delta(\mathbf{x-x'})\,\delta^{(a)}(\tau-\tau'),
\eeq
where the inhomogeneity consists of the delta function in space and the antiperiodic repetitive delta function in 
imaginary time
\beq\label{antiperiod-delta}
\delta^{(a)}(\tau-\tau') = \sum_{n=-\infty}^{\infty} (-1)^n\,\delta(\tau-\tau'+n\hbar\beta).
\eeq
The Gross-Pitaevskii equation (\ref{B-mot-eq-imag}) for the condensate and the inhomogeneous Schr{\"o}dinger equation 
(\ref{F-mot-eq-imag}) for fermions form a set of coupled equations in imaginary time. Therein the condensate 
wave function $\Psi_B(\mathbf{x},\tau)$ in the Gross-Pitaevskii equation (\ref{B-mot-eq-imag}) is modified by the 
fermionic 
Green function $G_F(\mathbf{x},\tau;\mathbf{x},\tau)$ and, vice versa, the condensate wave function $\Psi_B(\mathbf{x},
\tau)$ 
influences the fermionic Green function $G_F(\mathbf{x},\tau;\mathbf{x'},\tau')$ in the inhomogeneous Schr{\"o}dinger 
equation 
(\ref{F-mot-eq-imag}). A Wick rotation $\tau=i t$ and the omission of the chemical potentials in 
Eqs.~(\ref{B-mot-eq-imag}) 
and (\ref{F-mot-eq-imag}) leads to two coupled equations of motion in real time, which describe the dynamics 
in the Bose-Fermi mixture.
\subsection{Fermionic Green Function}\label{Green-function}
In the following we obtain an approximative solution for the
fermionic Green function $G_F({\mathbf x}, \tau ; {\mathbf x}', \tau')$.
To this end we restrict ourselves to a stationary BEC so that the imaginary
time-dependent Gross-Pitaevskii equation (\ref{B-mot-eq-imag}) reduces to
\beq\label{B-mot-eq-imag-ot}
\left[-\frac{\hbar^2}{2m_B}\Delta+V_B({\mathbf{x}})-\mu_B+g_{BB}
|\Psi_B(\mathbf{x})|^2 - g_{BF} G_F(\mathbf{x},\tau;\mathbf{x},\tau) \right]
\Psi_B(\mathbf{x}) = 0,
\eeq
and the effective fermionic Hamiltonian (\ref{Hamilt-F}) no longer depends explicitly
on the imaginary time $\tau$:
\beq\label{Hamilt-F-ot}
\hat H_{F,\mathrm{eff}}(\mathbf{x}) = -\frac{\hbar^2}{2m_F}\Delta+V_F(\mathbf{x})+
g_{BF}\,|\Psi_B(\mathbf{x})|^2.
\eeq
Now we assume that both the fermionic trap potential $V_F(\mathbf{x})$ and the condensate wave function
$\Psi_B(\mathbf{x})$ have only a weak
spatial dependence so that the semiclassical approximation can be
applied. Therein the eigenvalue problem
\beq
\hat H_{F,\mathrm{eff}}({\mathbf x})\,\psi_\mathbf{k} (\mathbf{x}) = E(\mathbf{k},\mathbf{x})\,\psi_\mathbf{k}(\mathbf{x})
\eeq
is approximately solved by plane waves
\beq
\psi_\mathbf{k}(\mathbf{x}) = \frac{e^{i \mathbf{k\cdot x}}}{(2 \pi)^{3/2}}
\eeq
and the semiclassical energy spectrum
\beq
E(\mathbf{k,x}) = \frac{\hbar^2\mathbf{k}^2}{2m_F}+V_F(\mathbf{x})+g_{BF} |\Psi_B(\mathbf{x})|^2.
\eeq
Thus, the semiclassical solution of Eq.~(\ref{F-mot-eq-imag}) yields for the fermionic Green
function
\begin{multline}\label{ferm-green-semicl}
G_F(\mathbf{x},\tau;\mathbf{x'},\tau') = \int \frac{d^3 k}{(2 \pi)^3} \, e^{i \, 
\mathbf{k}\cdot(\mathbf{x}-\mathbf{x'})}\\
\times \,\frac{\Theta(\tau - \tau')\,e^{-[E(\mathbf{k}, (\mathbf{x}+ \mathbf{x'})/2)-\mu_F](\tau - \tau'
- \hbar \beta /2)/ \hbar}-
\Theta(\tau' - \tau)\,e^{-[E(\mathbf{k}, (\mathbf{x}+\mathbf{x'})/2)-\mu_F](\tau -\tau'
+ \hbar \beta /2)/ \hbar}}{2 \cosh \beta
[E(\mathbf{k},(\mathbf{x}+\mathbf{x'})/2 ) -\mu_F]/2}  \, .
\end{multline}
The limit of equal imaginary times follows from $\tau ' \downarrow \tau$:
\beq
G_F(\mathbf{x},\tau ; {\mathbf x'},\tau)  = -\int \frac{d^3 k}{(2 \pi)^3}
\frac{e^{i \,\mathbf{k} \cdot ( \mathbf{x} - \mathbf{x'} )}}{e^{\beta \left[E({\mathbf k},
(\mathbf{x}+ \mathbf{x'})/ 2) - \mu_F\right]}+1}\,.
\eeq
Thus, the fermionic particle density $n_F(\mathbf{x}) = -G_F(\mathbf{x},\tau;\mathbf{x},\tau)$ is given by
\beq
\label{IINT}
n_F(\mathbf{x}) = \int \frac{d^3k}{(2\pi)^3}\, \frac{1}{e^{\beta\left[\hbar^2 \mathbf{k}^2/2m_F
-\tilde\mu_F(\mathbf{x})\right]}+1}
\eeq
with the local chemical potential
\beq
\label{ferm-dens}
\tilde\mu_F(\mathbf{x})=\mu_F-V_F(\mathbf{x})-g_{BF}\,|\Psi_B(\mathbf{x})|^2,
\eeq
which represents the kinetic energy of the fermion in the highest energetic state, when it is located at the \mbox{space point $\mathbf{x}$}. In order to evaluate the integral (\ref{IINT}) 
in momentum space, we apply spherical coordinates and the substitution $\varepsilon(k)=\hbar^2k^2/2m_F$:
\beq
\label{ferm-denss}
n_F(\mathbf{x}) = \frac{1}{\Gamma(3/2)}\,\left(\frac{m_F}{2\pi\hbar^2}\right)^{3/2} \int_0^\infty 
\frac{d\varepsilon\,\varepsilon^{1/2}}{e^{\beta
\left[\varepsilon-\tilde\mu_F(\mathbf{x})\right]}+1}.
\eeq
In the low-temperature limit $T\downarrow 0$, the Sommerfeld expansion \cite{Ashcroft} of Eq.~(\ref{ferm-denss}) 
yields in the lowest order
\beq\label{ferm-dens1}
n_F(\mathbf{x}) = \kappa\, \Theta\left(\tilde\mu_F(\mathbf{x})\right)\,\tilde\mu_F^{3/2}(\mathbf{x})
\eeq
with the abbreviation $\kappa = (2m_F)^{3/2}/6\pi^2\hbar^3$. Finally, we insert the result for the fermionic particle 
density (\ref{ferm-dens1}) into the Gross-Pitaevskii equation (\ref{B-mot-eq-imag-ot}): 
\beq
\label{loc-GP-eq}
\left[-\frac{\hbar^2}{2m_B}\Delta+V_B({\mathbf{x}})-\mu_B+g_{BB} |\Psi_B(\mathbf{x})|^2 + \kappa g_{BF}
\, \Theta\left(\tilde\mu_F(\mathbf{x})\right)\,\tilde\mu_F^{3/2}(\mathbf{x}) \right]\Psi_B(\mathbf{x}) = 0.
\eeq
Note that this stationary Gross-Pitaevskii equation (\ref{loc-GP-eq}) follows from extremizing the effective action
\beq
\label{F9}
\Gamma [\Psi_B^*,\Psi_B] = \int d^3x\,\left\{ \Psi_B^*(\mathbf{x})\left[-\frac{\hbar^2}{2m_B}\Delta
+V_B({\mathbf{x}})-\mu_B +\frac{g_{BB}}{2}
\,|\Psi_B(\mathbf{x})|^2\right] \Psi_B(\mathbf{x})
-\frac{2}{5}\, \kappa \, \Theta(\tilde\mu_F(\mathbf{x}))\, \tilde\mu_F^{5/2}(\mathbf{x}) \right\} 
\eeq
with respect to $\Psi_B^*(\mathbf{x})$. This effective action can also be obtained from Eqs.~(\ref{A-B}) and 
(\ref{F8})--(\ref{Hamilt-F}) by evaluating the fermionic functional integral (\ref{Z-F}) in the semiclassical 
approximation for a stationary BEC and by performing the Sommerfeld expansion in the low-temperature limit.  
Identifying the extremum of the effective action $\Gamma$ with the grand-canonical free energy ${\cal F}$,
the number of bosons and fermions are obtained from (\ref{F9}) via the normalization condition
\beq\label{normaliz}
N_j = -\frac{\partial \mathcal{F}}{\partial \mu_j} = \int d^Dx\,n_j(\mathbf{x}), \qquad\qquad j = B,F,
\eeq
where the particle densities of bosons and fermions read
\beq\label{n-B}
n_B(\mathbf{x}) &=& |\Psi_B(\mathbf{x})|^2,
\\ \label{n-F}
n_F(\mathbf{x}) &=& \kappa\, \Theta\left(\tilde\mu_F(\mathbf{x})\right)\, \tilde\mu_F^{3/2}(\mathbf{x}).
\eeq
\section{Density profiles}\label{sol-GP}
In this section we solve the stationary Gross-Pitaevskii equation (\ref{loc-GP-eq}) in the Thomas-Fermi 
approximation in order to calculate the boson and the fermion density distribution for the parameters of the 
experiment in Hamburg \cite{Sengstock}. There the $^{87}$Rb--$^{40}$K boson-fermion mixture is 
confined in a three-dimensional rotationally symmetric harmonic trap
\beq\label{anisotr-oszill}
V_i(\mathbf{x}) = \frac{m_i}{2} \left(\omega_{i,r}^2r^2+\omega_{i,z}^2z^2\right),
\qquad\qquad
i = B,F,
\eeq
where we use cylindrical coordinates $\{r,\phi,z\}$. The frequencies for 
both species are related by $\omega_{F,k}= \sqrt{m_B/m_F}\,\omega_{B,k}$ for $k=r,z$, so that $m_B\,\omega_{B,k}^2/2 = m_F\,\omega_{F,k}^2/2$.
\subsection{Thomas-Fermi approximation}
Assuming that the potential and interaction energy are larger than the kinetic energy,
we can use the Thomas-Fermi approximation, where the kinetic term 
in the Gross-Pitaevskii equation can be neglected. With this approximation the Gross-Pitaevskii equation 
(\ref{loc-GP-eq}) reduces with the help of Eqs.~(\ref{n-B}) and (\ref{n-F}) to an algebraic equation with respect 
to the bosonic particle density $n_B(\mathbf{x})$ \cite{Molmer}:
\beq\label{algebr-GP-eq}
V_B({\mathbf{x}})-\mu_B+g_{BB}\,n_B(\mathbf{x}) +g_{BF}\,n_F(\mathbf{x}) = 0.
\eeq
The last term in Eq.~(\ref{algebr-GP-eq}) stands for the fermionic particle density
\beq\label{dens-F}
n_F(\mathbf{x}) = \kappa\,\Theta\left(\mu_F-V_F(\mathbf{x})-g_{BF}\,n_B(\mathbf{x})\right) 
\left[\mu_F-V_F(\mathbf{x})-g_{BF}
\,n_B(\mathbf{x})\right]^{3/2},
\eeq
which modulates the bosonic density profile $n_B(\mathbf{x})$ and vice versa.
\subsection{Vanishing Boson-Fermion Interaction}
Now we discuss the special case of vanishing boson-fermion interaction $g_{BF} \to 0$, where we mark all quantities 
with the upper index $(0)$. On the one hand, Eq.~(\ref{algebr-GP-eq}) reduces then to the well-known particle 
density of a pure BEC in the Thomas-Fermi approximation
\beq\label{th-fermi-dens}
n_B^{(0)}(\mathbf{x}) = \frac{1}{g_{BB}}\,\Theta\left(\mu_B^{(0)}-V_B(\mathbf{x})\right)\left[\mu_B^{(0)}
-V_B(\mathbf{x})\right],
\eeq
provided that $g_{BB} > 0$. In case of an attractive interacting BEC with negative $g_{BB}$, the third term in 
Eq.~(\ref{algebr-GP-eq}) cannot be balanced in the trap center at $\mathbf{x=0}$ by the remaining chemical 
potential $\mu_B^{(0)}$ leading to a collapse of the BEC. The particle density (\ref{dens-F}) of fermions, on the 
other hand, becomes independent of that of bosons:
\beq\label{dens-F1}
n_F^{(0)}(\mathbf{x}) = \kappa\,\Theta\left(\mu_F^{(0)}-V_F(\mathbf{x})\right) \left[\mu_F^{(0)}
-V_F(\mathbf{x})\right]^{3/2}.
\eeq
Thus, the clouds of the BEC and the Fermi gas coexist undisturbed from each other.
Setting the particle densities (\ref{th-fermi-dens}) and (\ref{dens-F1}) zero yields the Thomas-Fermi-radii of the BEC 
and the Fermi cloud along the respective axes:
\beq\label{Th-Fermi-rad}
R_{i,k}^{(0)} = \sqrt{\frac{2\mu_i^{(0)}}{m_i\, \omega_{i,k}^2}},
\qquad\qquad i = B,F.
\eeq
The chemical potentials $\mu_B^{(0)}$ and $\mu_F^{(0)}$ are determined by the particle numbers $N_B$ and $N_F$ via the 
normalization (\ref{normaliz}):
\beq\label{mueb}
\mu_B^{(0)} = \left(\frac{15a_{BB}N_B}{\tilde L_B}\right)^{2/5} \frac{\hbar\tilde\omega_B}{2},
\qquad\qquad
\mu_F^{(0)}= (6N_F)^{1/3}\,\hbar\tilde\omega_F,
\eeq
where $\tilde L_i = \sqrt{\hbar/m_i\,\tilde\omega_i}$ denotes the geometrical  average of the oscillator lengths $L_{i,k} 
= \sqrt{\hbar/m_i\,\omega_{i,k}}$. Inserting the chemical potentials (\ref{mueb}) into the respective particle densities 
(\ref{th-fermi-dens}) and (\ref{dens-F1}), the maximum of the latter in the trap center results in
\beq\label{th-f-dens-b-max}
n_B^{(0)}(\mathbf{0}) = \frac{(15N_B)^{2/5}}{8\pi a_{BB}^{3/5}\,\tilde L_B^{12/5}},
\qquad\qquad
n_F^{(0)}(\mathbf{0}) = \frac{2 N_F^{1/2}}{3^{1/2}\pi^2\,\tilde L_F^3}.
\eeq
In the Thomas-Fermi approximation the BEC needs a non-vanishing repulsion between the bosons in order to prevent the 
density from becoming infinite or complex, which indicates a collapse of the BEC, whereas the quantum pressure in 
the noninteracting Fermi gas preserves the latter from a collapse. The maximum fermionic particle density increases 
faster with the particle number than the maximum  bosonic particle density, whereas the latter depends not only on 
the oscillator length but also on $a_{BB}$, the parameter for the bosonic interaction strength. The particle 
densities of both species in an undisturbed BEC and Fermi gas are plotted for typical particle numbers $N_B$ and $N_F$ 
as thin lines in Figure \ref{fig:B-F-dens}.
\subsection{Non-Vanishing Boson-Fermion Interaction}\label{discus-dens}
\begin{figure}[t]
\centerline{\includegraphics[scale=1.0]{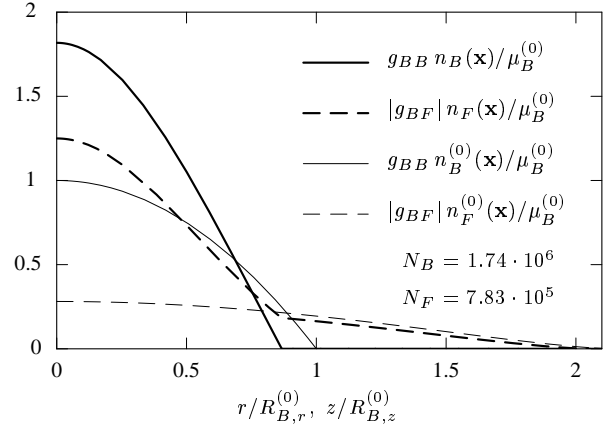}}
\caption{Comparison of the dimensionless particle densities for bosons and for fermions between a disturbed mixture 
(thick lines) and an undisturbed BEC and Fermi gas  (thin lines). The densities are plotted versus the coordinates 
$r$ at the plane $z=0$ and $z$ at the line $r=0$, respectively, in units of the Thomas-Fermi radii of the undisturbed 
BEC. The densities of the disturbed mixture are the solution of Eqs.~(\ref{algebr-GP-eq}) and (\ref{dens-F}), whereas 
those of an undisturbed BEC and Fermi gas are given by Eqs.~(\ref{th-fermi-dens}) and (\ref{dens-F1}), respectively, 
for an example of typical  particle numbers $N_B$, $N_F$ of the Hamburg experiment \cite{Sengstock}.}
\label{fig:B-F-dens}
\end{figure}
After having treated the case $g_{BF}=0$ in the previous section, 
we now discuss the algebraic equation (\ref{algebr-GP-eq}) for a 
BEC which interacts with fermions. In order to plot the corresponding bosonic and fermionic densities in 
Figure~\ref{fig:B-F-dens}, we insert the fermionic density (\ref{dens-F}) 
into the algebraic equation (\ref{algebr-GP-eq}) and obtain:
\beq
\label{algebr-eq1}
V_B({\mathbf{x}})-\mu_B+g_{BB}\,n_B(\mathbf{x}) = -g_{BF}\,\kappa \left[\mu_F-V_F(\mathbf{x})
-g_{BF}\,n_B(\mathbf{x})\right]^{3/2}.
\eeq
Squaring (\ref{algebr-eq1}) leads to a cubic equation with respect to the condensate density $n_B(\mathbf{x})$:
\begin{multline}\label{cubic-eq}
n_B^3(\mathbf{x})+\left\{\frac{g_{BB}^2}{\kappa^2\,g_{BF}^5}-\frac{3[\mu_F-V_F(\mathbf{x})]}{g_{BF}}\right\}
n_B^2(\mathbf{x})
+\left\{\frac{2g_{BB}[V_B(\mathbf{x})-\mu_B]}{\kappa^2\,g_{BF}^5}+\frac{3[\mu_F-V_F(\mathbf{x})]^2}{g_{BF}^2}\right\}
n_B(\mathbf{x})\\
+\frac{[V_B(\mathbf{x})-\mu_B]^2}{\kappa^2\,g_{BF}^5}-\frac{[\mu_F-V_F(\mathbf{x})]^3}{g_{BF}^3} = 0,
\end{multline}
It is of the form
\beq\label{cub-eq}
x^3+ax^2+bx+c = 0
\eeq
with real coefficients $a,b,c$ and its three solutions $x_1,x_2,x_3$ can be found with the help of 
the Cardanian formula \cite{Greiner2}. The first solution $x_1$ remains always real, whereas the other solutions $x_2$ and $x_3$ 
are either real for 
\beq\label{compl-cond}
\left[-\frac{a^2}{9}+\frac{b}{3}\right]^3\leq -\left[-\frac{a^3}{27}+\frac{ab}{6}-\frac{c}{2}\right]^2
\eeq
or conjugate complex otherwise. Although all three solutions obey the cubic equation (\ref{cub-eq}), i.e.~Eq.~(\ref{cubic-eq}), only the last 
solution $x_3$ turns out to satisfy the algebraic equation (\ref{algebr-eq1}). Inserting the two other solutions 
into the algebraic equation yields the right value but opposite signs on both sides of the algebraic equation 
(\ref{algebr-eq1}). The reason is that Eq.~(\ref{algebr-eq1}) represents the root of the cubic equation 
(\ref{cubic-eq}), therefore an equation of the order $3/2$, which reduces the number of possible solutions.

The resulting particle densities for both species are plotted as thick lines in Figure \ref{fig:B-F-dens} for typical 
particle numbers $N_B$ and $N_F$ using the experimental parameters in Table \ref{tab:exp-parameter}. In order to 
determine the solution outside the BEC, especially the particle density of fermions, we must not forget the origin 
of the algebraic equation (\ref{algebr-GP-eq}) from the Gross-Pitaevskii equation (\ref{loc-GP-eq}). Although the 
solution $n_B(\mathbf{x})$ becomes negative outside the BEC, we have to set it equal zero there, as the density 
must be positive by definition. Thus, the condensate wave function $\Psi_B(\mathbf{x})$ becomes also zero outside the 
BEC and fulfills the Gross-Pitaevskii equation (\ref{loc-GP-eq}) in a trivial way regardless of the expression 
inside the brackets, which is, apart from the neglected kinetic energy term, the algebraic equation (\ref{algebr-GP-eq}). 
Hence, the fermionic particle density outside the BEC is described by Eq.~(\ref{dens-F}) with $n_B(\mathbf{x})=0$ 
without obeying the algebraic equation (\ref{algebr-GP-eq}).

The particle density of the BEC in the Thomas-Fermi approximation can be written in the following form:
\beq\label{dens-B-impl}
n_B(\mathbf{x}) = \frac{1}{g_{BB}}\,\left[\mu_B-V_B({\mathbf{x}}) -g_{BF}\,n_F(\mathbf{x})\right] \, .
\eeq
From the fermionic particle density (\ref{dens-F}) we deduce that its maximum occurs in the trap center 
$\mathbf{x=0}$ due to the negative $g_{BF}$. Thus, the BEC possesses its largest density also in the trap center. 
In other words, the particle densities of both species intensify each other in the overlapping region due to the 
strong attraction between bosons and fermions as is shown in Figure \ref{fig:B-F-dens}. With increasing distance 
from the trap center both densities (\ref{dens-F}) and (\ref{dens-B-impl}) decrease quickly within their overlap 
due to the interspecies interaction terms and due to the increasing trap potentials, whereas outside of 
the overlap only the latter reason is responsible for their decreasing. This behavior is shown in Figure 
\ref{fig:B-F-dens} for the Hamburg experiment. Which of both clouds has the larger extension depends on the 
particle numbers $N_B$ and $N_F$ and therefore on the chemical potentials $\mu_B$ and $\mu_F$. Usually, the BEC is 
surrounded by the Fermi gas unless $N_B\gg N_F$. Both chemical potentials, which represent the total energy of a 
particle of the corresponding species, are smaller compared with $\mu_B^{(0)}$ and $\mu_F^{(0)}$ of the undisturbed 
BEC and Fermi gas, since the particles possess, besides the neglected kinetic energy, the potential energy due to 
the trap and the intraspecies interaction energy an additional negative interaction 
energy due to the interspecies interaction. 
Figure \ref{fig:energy} shows these energies of a boson in units of $V_B(R_{B,r},0)
=V_B(0,R_{B,z})=\mu_B^{(0)}$ for the Hamburg experiment, where we see that $\mu_B \approx 0.6\, \mu_B^{(0)}$. These reduced 
chemical 
potentials lead to decreasing Thomas-Fermi radii for both species. Indeed, one can see in Figure \ref{fig:B-F-dens} 
a reduction of the Thomas-Fermi radii of the BEC to $R_{B,k}\approx 0.8\,R_{B,k}^{(0)}$. Hence the attractive interaction 
between both species leads to an additional confinement of both the BEC and the Fermi gas within their overlap.

A direct quantitative comparison with the density of the experimental probe is not possible since a measurement of the 
density in the trap fails due to the smallness of the probe. Only after the expansion of the probe, when the trap 
potential was suddenly switched off, its density can be measured taking the absorption image of the optical density of $^{87}$Rb and $^{40}$K. This would allow to compare the densities only qualitatively, since the density in the probe 
is changed due to both the ballistic expansion and the dynamics in the probe.
\subsection{Validity of Thomas-Fermi approximation}
\begin{figure}[t]
\centerline{\includegraphics[scale=1.0]{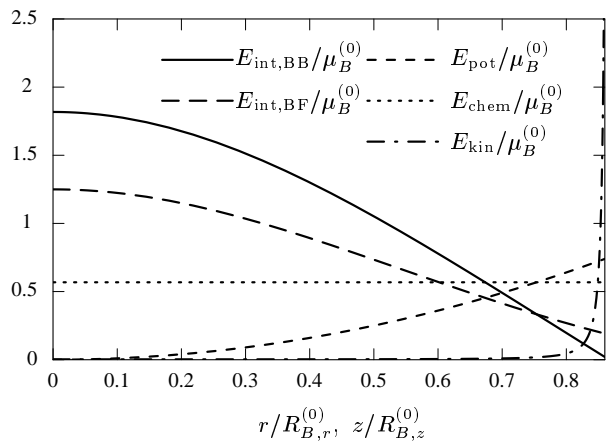}}
\caption{Comparison of the kinetic energy 
$E_\mathrm{kin}=-\hbar^2\Delta \Psi_B(\mathbf{x})/(2 m_B\,\Psi_B(\mathbf{x}))$ of a boson, with its remaining energies, 
namely the intraspecies interaction energy $E_\mathrm{int,BB}=g_{BB}\,n_B(\mathbf{x})$, the interspecies 
interaction energy $E_\mathrm{int,BF}=|g_{BF}|\,n_F(\mathbf{x})$, the potential energy 
$E_\mathrm{pot}=V_B(\mathbf{x})$ due to the trap and the chemical potential $E_\mathrm{chem}=\mu_B$. 
All energies are related to the value $\mu_B^{(0)}$ of the potential energy at the  boundary of the 
undisturbed BEC cloud and are plotted versus the coordinates $r$ and $z$ within the BEC cloud for the situation 
in Figure \ref{fig:B-F-dens}.}
\label{fig:energy}
\end{figure}
In order to check the validity of the Thomas-Fermi approximation, we have plotted in Figure \ref{fig:energy} the 
kinetic energy of a boson
\beq\label{kin-energ-b}
E_\mathrm{kin} = \frac{-\hbar^2\Delta \Psi_B(\mathbf{x})}{2 m_B \,\Psi_B(\mathbf{x})}
\eeq
in units of the value $\mu_B^{(0)}$ of the potential energy at the boundary of the undisturbed BEC. The kinetic 
energy is, indeed, negligible in a wide bulk range from the trap center to just before the boundary of the disturbed 
BEC. Thus, the Thomas-Fermi approximation gives very accurate results except in the outermost 10\% of the Thomas-Fermi 
radii. The kinetic energy diverges at the BEC boundary because the condensate wave function $\Psi_B(\mathbf{x})$, which 
represents the square root of the particle density $n_B(\mathbf{x})$, occurs due to the Laplacian derivative in 
Eq.~(\ref{kin-energ-b}) in the denominator, which becomes zero at the BEC boundary. It is obvious that the proper 
solution of the Gross-Pitaevskii equation (\ref{loc-GP-eq}) would match the one of the algebraic equation 
(\ref{algebr-GP-eq}) from the trap center to just before the BEC boundary, where it tends smoothly to zero, thus 
improving the sharp bend in the graph of the fermionic particle density there to be smooth.
\begin{figure}[t]
\centerline{\includegraphics[scale=1.0]{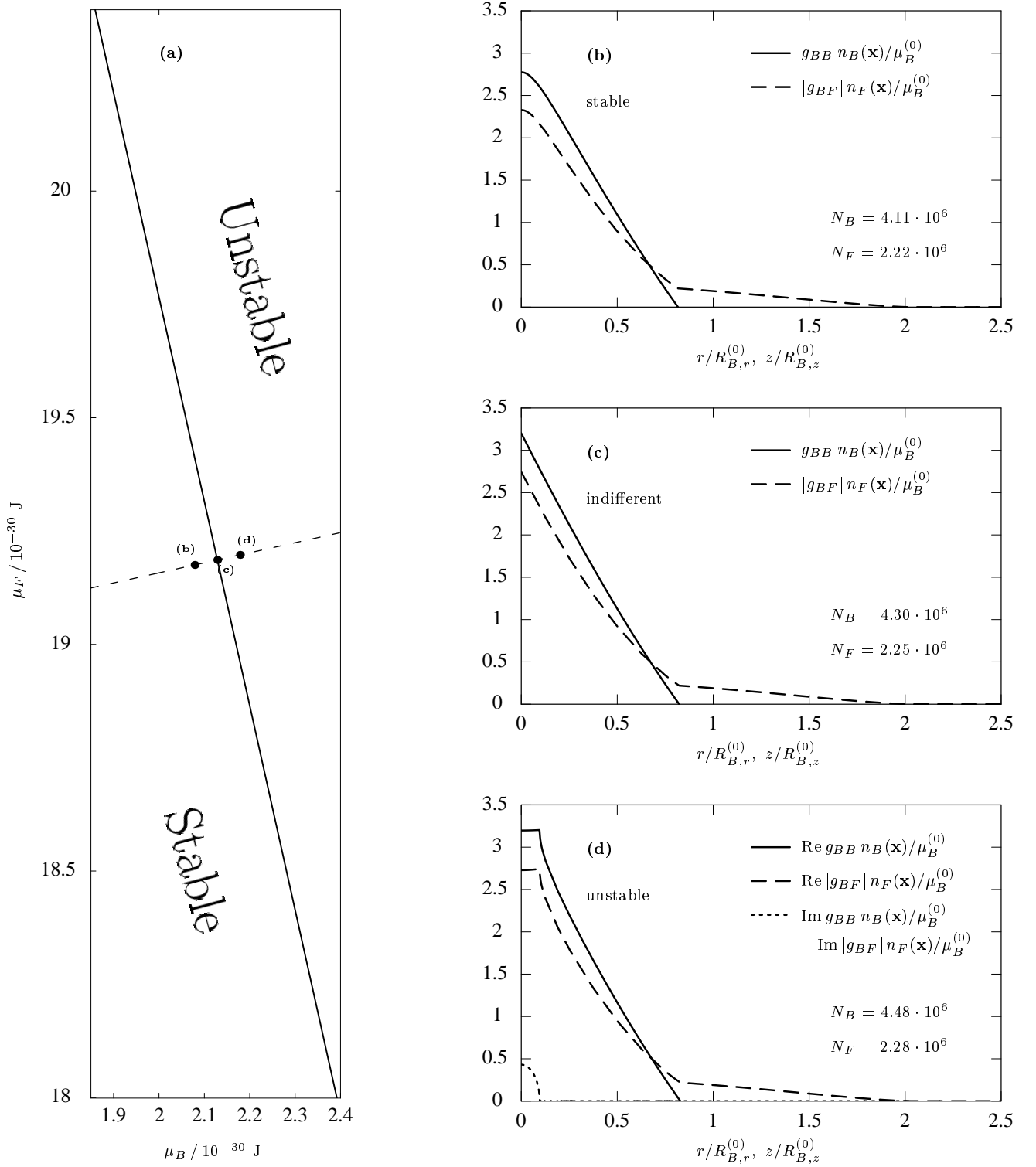}}
\caption{(a) Stability diagram with respect to the chemical potentials. The solid line separates the stable 
region on the left from the unstable region on the right according to Eq.~(\ref{crit-line-mu}). 
The pictures (b)--(d) show a sequence how the bosonic 
and fermionic particle densities versus the coordinates $r$ at the plane $z=0$ and $z$ at the line $r=0$, 
respectively, change on the road from stability to instability. The $(\mu_B,\mu_F)$-pairs belonging to (b)--(d) 
are located on the dashed line in (a) and are equally spaced $\Delta\mu_B=5\cdot 10^{-32}$ J apart giving rise to 
particle 
number differences of about $\Delta N_B\approx 1.8\cdot 10^5$ and $\Delta N_F\approx 3\cdot 10^4$, respectively.}
\label{fig:B-F-dens-beside-coll}
\end{figure}
%
%
%
\subsection{Complex solutions}\label{compl-sol}
As already discussed in Subsection \ref{discus-dens}, the solution $n_B(\mathbf{x})$ of the algebraic equation 
(\ref{algebr-GP-eq}) does not always remain real by varying 
the chemical potentials $\mu_B$ and $\mu_F$. From the condition (\ref{compl-cond}) we extract 
a straight line in the \mbox{$\mu_B,\mu_F$-plane}
\beq
\label{crit-line-mu}
\mu_F = \frac{4g_{BB}^2}{27\kappa^2\,g_{BF}^4}+\frac{g_{BF}}{g_{BB}}\,\mu_B,
\eeq
which separates the half plane with a complex solution from that with a real solution. This situation is depicted for 
the Hamburg experiment in Figure \ref{fig:B-F-dens-beside-coll} (a). Due to the results of the subsequent 
Section \ref{Stability-collaps}, we can associate the complex solution to an instability of the mixture with respect 
to collapse. Figures \ref{fig:B-F-dens-beside-coll} (b)--(d) demonstrate the change of the density 
profiles $n_B(\mathbf{x})$ and $n_F(\mathbf{x})$ on the road from stability to instability by increasing the particle 
numbers with $\Delta N_B\approx 1.8\cdot 10^5$ and $\Delta N_F\approx 3\cdot 10^4$ each. The respective pairs 
$(\mu_B,\mu_F)$ lie on the dashed line, 
which is arranged perpendicular to the critical solid 
line in Figure \ref{fig:B-F-dens-beside-coll} (a), so for Figure \ref{fig:B-F-dens-beside-coll} (b) 
just below the critical line, 
for Figure \ref{fig:B-F-dens-beside-coll} (c) on, and for Figure \ref{fig:B-F-dens-beside-coll} (d) 
just above. Figure \ref{fig:B-F-dens-beside-coll} (b) shows a stable configuration, 
Figure \ref{fig:B-F-dens-beside-coll} (c) is on the boundary 
between stable and unstable where the densities pile up at the trap center to a peak, and 
in Figure \ref{fig:B-F-dens-beside-coll} (d) the real part 
of the densities is chopped off at the trap center, so that there an imaginary part of both densities occurs. This 
imaginary part starts to appear simultaneously for both components at the density maximum $\mathbf{x=0}$ and grows 
in magnitude and in extension out of the trap center with increasing $N_B$ and $N_F$. Such an anomaly in the densities 
was observed in the Hamburg experiment, where the evolution of an overcritical mixture is shown in Figure 1 of 
Ref.~\cite{Sengstock}. 
\section{Stability against collapse}\label{Stability-collaps}
Here we determine the stability border both within the Thomas-Fermi approximation and, in a separate variational 
calculation, beyond the Thomas-Fermi approximation. The stability border turns out to depend strongly on the value 
of the interspecies s-wave scattering length. Therefore, comparing our theoretical results with the experimental 
measurements allows to extract a trustworthy value for this crucial s-wave scattering length.
\subsection{Thomas-Fermi Approximation}\label{Stab-T-F}
\begin{figure}[t]
\centerline{\includegraphics[scale=1.0]{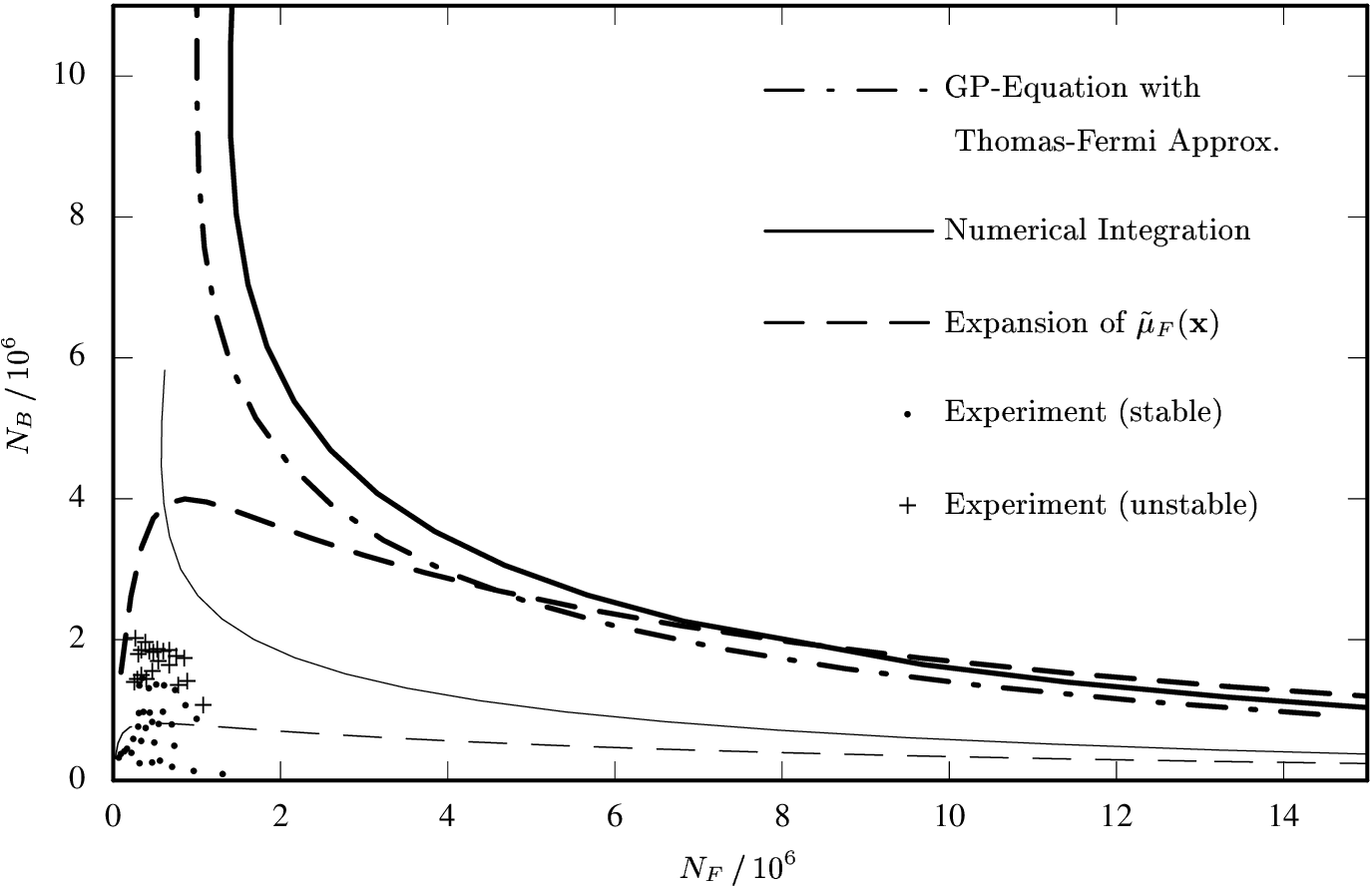}}
\caption{Stability diagram for the $^{87}$Rb--$^{40}$K mixture of the Hamburg experiment. The particle number 
pairs $(N_B,N_F)$ below a certain line belong to a stable mixture wheras those above the line indicate an unstable 
mixture tending to collapse. The thin lines correspond to the quantum mechanical limit with the ratio 
$\lambda_\mathrm{QM}=\omega_z/\omega_r$ and the thick lines represent the Thomas-Fermi limit with the ratio 
$\lambda_\mathrm{TF}=(\omega_z/\omega_r)^2$. 
The points are obtained in the experiment by analyzing decay series in various particle number regimes and are 
assigned to stable mixtures (dots) and to unstable mixtures (crosses) \cite{Sengstock}.}
\label{fig:stab-diag-h}
\end{figure}

\begin{figure}[t]
\centerline{\includegraphics[scale=1.0]{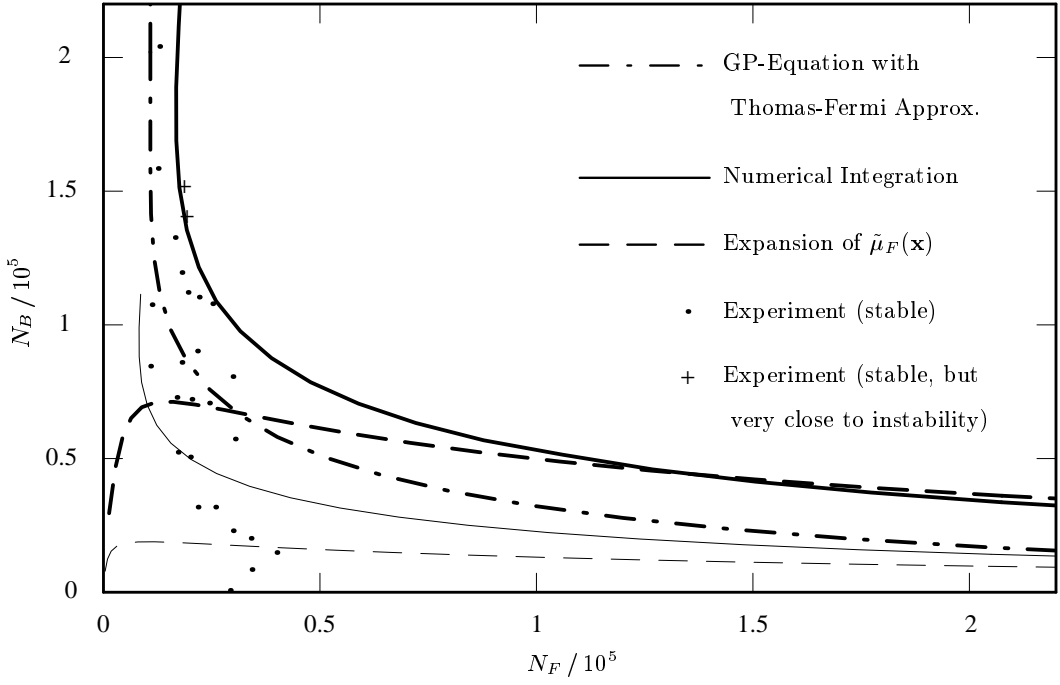}}
\caption{Stability diagram for the $^{87}$Rb--$^{40}$K mixture of the Florence experiment. The particle number pairs 
$(N_B,N_F)$ below a certain line belong to a stable mixture whereas those above the line indicate an unstable mixture 
tending to collapse. The thin lines correspond to the quantum mechanical limit with the ratio 
$\lambda_\mathrm{QM}=\omega_z/\omega_r$ 
and the thick lines represent the Thomas-Fermi limit with the ratio $\lambda_\mathrm{TF}=(\omega_z/\omega_r)^2$. 
The crosses represent 
mixtures in the experiment which are found very close to the instability, whereas the dots indicate stable mixtures 
\cite{Ferlaino}.}
\label{fig:stab-diag-f}
\end{figure}

In Subsection \ref{compl-sol} we described in detail the behavior of the particle densities $n_B(\mathbf{x})$ and 
$n_F(\mathbf{x})$ as the solution of the stationary Gross-Pitaevskii equation within the Thomas-Fermi approximation 
for varying particle numbers $N_B$ and $N_F$. We assigned the emergence of a complex density to a loss of the stability 
against collapse and found a border in form of a line in the $(\mu_B,\mu_F)$-plane, which separates the stable and 
unstable 
regions as shown in Figure \ref{fig:B-F-dens-beside-coll} (a). In order to obtain a stability diagram in the 
$(N_B,N_F)$-plane, we have evaluated the corresponding particle numbers from the chemical potentials by integrating 
out the respective particle densities according to the normalization condition (\ref{normaliz}). The result 
for the Hamburg and the Florence experiment is given 
by the dot-dashed line in Figures \ref{fig:stab-diag-h} and \ref{fig:stab-diag-f}, respectively. 
Mixtures with particle number pairs $(N_B,N_F)$ below this line are stable whereas particle 
number pairs above the line indicate an unstable mixture tending to collapse. The critical particle numbers of both 
species behave, roughly spoken, inversely proportional to each other in a wide range. Whereas the critical number of 
bosons $N_{B\mathrm{crit}}$ tends to zero when the number of fermions increases, the critical number of fermions 
$N_{F\mathrm{crit}}$ remains finite and constant when the number of bosons is enlarged. This situation, where the 
line in the stability diagram becomes vertical, happens when the BEC cloud becomes so large that it surrounds the 
Fermi gas. This defines a minimal number of fermions $N_{F,{\rm min}}$ below which the mixture remains stable
irrespective of the number of bosons $N_B$. In order to estimate $N_{F,{\rm min}}$, we have to integrate the
fermionic particle density (\ref{dens-F}) by taking into account the stability condition 
(\ref{crit-line-mu}) when solving
the algebraic equation (\ref{algebr-eq1}) for the condensate density $n_B(\mathbf{x})$. In this way we find
that the fermionic particle density (\ref{dens-F}) does not explicitly depend on both chemical potentials
$\mu_F$ and $\mu_B$. A numerical evaluation for the Hamburg (Florence) experiment yields 
$N_{F,{\rm min}}=9,99 \,\cdot 10^5$ ($N_{F,{\rm min}}=1,08 \,\cdot 10^4$) in accordance with the
dot-dashed instability line shown in Figure \ref{fig:stab-diag-h} (Figure \ref{fig:stab-diag-f}).
\subsection{Variational Method}\label{variation}
\begin{figure}[t]
\begin{minipage}[t]{0.48\textwidth}
\begin{picture}(100,60)
\centerline{\includegraphics[scale=1.0]{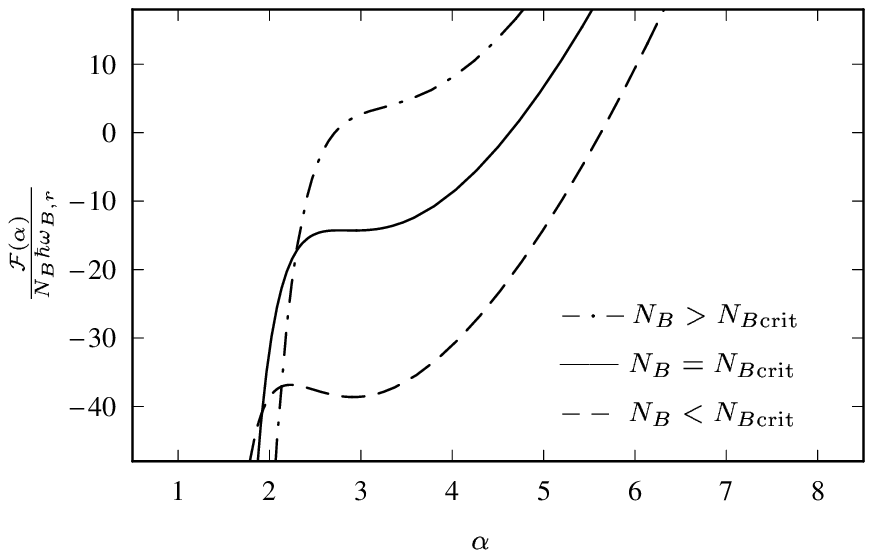}}
\end{picture}
\caption{Grand-canonical free energy $\mathcal{F}(\alpha)/N_B\hbar\omega_{B,r}$ versus the variational parameter 
$\alpha$ for various 
boson numbers $N_B$. A local minimum indicates a metastable state of the mixture.}
\label{fig:F-w-depend}
\end{minipage}\hfill
\begin{minipage}[t]{0.48\textwidth}
\begin{picture}(100,60)
\centerline{\includegraphics[scale=1.0]{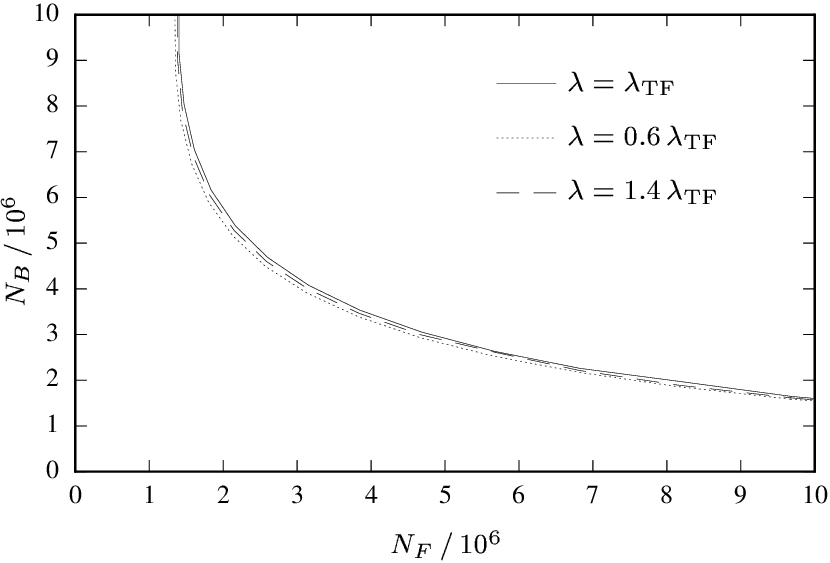}}
\end{picture}
\caption{Comparison of two stability borders whose ratio $\lambda$ is 40\% smaller and larger, respectively, than 
$\lambda_\mathrm{TF}$ in order to estimate the proper ratio $\lambda$ which allows the largest possible particle number 
for a stable 
condensate of the Hamburg experiment.}
\label{fig:30-ab-bel}
\end{minipage}
\end{figure}

Another approach to determine the stability border for a $^{87}$Rb--$^{40}$K mixture is based on extremizing the 
grand-canonical free energy (\ref{F9}) with the local chemical potential (\ref{ferm-dens}).
In contrast to the Thomas-Fermi approximation in Subsection \ref{Stab-T-F}, this approach takes into account the 
kinetic energy of the condensed atoms. But instead of varying the condensate wave function $\Psi_B(\mathbf{x})$ in the 
grand-canonical free energy (\ref{F9}), which leads to the Gross-Pitaevskii equation as described in Section 
\ref{Green-function}, we use the ground-state wave function of a three-dimensional anisotropic harmonic oscillator 
\beq\label{test-func}
\Psi_B(\mathbf{x}) = \sqrt{\frac{N_B}{\pi^{3/2}\alpha^3\tilde L_B^3}}\,
\exp\left\{-\sum_{k=1}^{3}\frac{x_k^2}{2\alpha^2L_{B,k}^2}\right\}
\eeq
as a test function with  variational widths $\alpha\,L_{B,k}$.
The dimensionless factor $\alpha$, which scales the oscillator 
lengths of the Gaussian function (\ref{test-func}), serves as a variational parameter, which extremizes the free 
energy (\ref{F9}). The test function is normalized to $N_B$ bosons and obeys for $\alpha=1$ the Gross-Pitaevskii equation 
for a trapped noninteracting BEC:
\beq\label{harm-oscil}
\left[-\frac{\hbar^2}{2m_B}\Delta+V_B({\mathbf{x}})-\mu_B \right]\Psi_B(\mathbf{x}) = 0.
\eeq
Here we assume that the condensate wave  function $\Psi_B(\mathbf{x})$ has, also in case of intraspecies and interspecies 
two-particle interactions, qualitatively the shape of a Gaussian curve, as one can read off from Figure 
\ref{fig:B-F-dens}. Using cylindrical coordinates $\{r,\phi,z\}$, the test function (\ref{test-func}) reads
\beq\label{test-func2}
\Psi_B(\mathbf{x}) = \sqrt{\frac{N_B\lambda^{1/2}}{\pi^{3/2}\alpha^3 L_{B,r}^3}}\,
\exp\left\{-\frac{r^2+\lambda z^2}{2\alpha^2L_{B,r}^2}\right\}.
\eeq
Beside a uniform variation of the widths $\alpha L_{B,k}$ by the factor $\alpha$, we have to consider that the ratio 
$\lambda = (L_{B,z}/L_{B,r})^2$ could also change due to the interactions. In order to include this, we perform the 
calculation 
with two different ratios. On the one hand we use $\lambda_\mathrm{QM}=\omega_z/\omega_r$ which stands 
for the limit of vanishing interactions $g_{BB}\to 0$ and $g_{BF}\to 0$ and reflects the proper ratio of the oscillator 
lengths in the ground-state wave function of the quantum-mechanical harmonic oscillator (\ref{harm-oscil}). 
On the other hand, we set $\lambda_\mathrm{TF}=(\omega_z/\omega_r)^2$, which represents the Thomas-Fermi limit of 
negligible kinetic 
energy due to strong intraspecies and interspecies interaction. 
Inserting the test function (\ref{test-func2}) into the grand canonical free energy (\ref{F9}) reduces the latter 
from the functional $\mathcal{F}[\Psi_B^*,\Psi_B]$ to a function $\mathcal{F}(\alpha)$ of the parameter $\alpha$. 
As the test 
function (\ref{test-func2}) is normalized independent of $\mu_B$, the latter plays no longer a role in 
$\mathcal{F}(\alpha)$ 
and shifts the free energy only by a constant value. On the other hand, the fermionic chemical potential $\mu_F$ is 
needed for evaluating the respective fermion number $N_F$ by integrating out the fermionic particle density 
(\ref{dens-F}) according to the normalization (\ref{normaliz}). The dependence of the grand-canonical free energy 
$\mathcal{F}(\alpha)$ on the variational parameter $\alpha$ for given chemical potentials 
$\mu_B$ and $\mu_F$ is shown in Figure 
\ref{fig:F-w-depend} for several boson numbers $N_B$. For $N_B<N_{B\mathrm{crit}}$ the free energy $\mathcal{F}(\alpha)$ 
possesses a local minimum which corresponds to a metastable state of the mixture. The condensate wave function 
$\Psi_B(\mathbf{x})$ has finite equilibrium widths $\alpha_\mathrm{eq} L_{B,k}$, where $\alpha_\mathrm{eq}$ denotes the 
parameter 
at the local minimum. When the boson number exceeds the critical value, i.e.~$N_B>N_{B\mathrm{crit}}$, the local 
minimum disappears so that the widths tend to zero in order to minimize $\mathcal{F}(\alpha)$. Just this happens when the 
mixture collapses. Thus, the border between stability and instability is given by the condition
\beq\label{stab-cond}
N_B = N_{B\rm{crit}}
\qquad\Leftrightarrow\qquad
\left.\frac{d\mathcal{F}(\alpha)}{d\alpha}\right|_{\alpha=\alpha_{\rm{crit}}} = \left.
\frac{d^2\mathcal{F}(\alpha)}{d\alpha^2}
\right|_{\alpha=\alpha_{\rm{crit}}} = 0,
\eeq
i.e. $\mathcal{F}(\alpha)$ has a point of inflexion at $\alpha=\alpha_{\rm{crit}}$. The appearance of the local 
minimum arises 
from the competition between the positive first three terms of the grand-canonical free energy (\ref{F9}) and the 
negative last term describing the influence of the fermions. We have determined the stability border within the 
variational method in two different ways, which we discuss in the following subsections.
\subsection{Numerical Integration}\label{num-int}
The most reliable approach is based on performing the integration in Eq.~(\ref{F9}) numerically. We have evaluated 
the critical boson numbers $N_{B\rm{crit}}$ for different values of $\mu_F$ in an iterative way until the condition 
(\ref{stab-cond}) is achieved with sufficient accuracy. The result is given by solid lines in the stability diagrams 
for the Hamburg and Florence experiment as shown in Figures \ref{fig:stab-diag-h} and \ref{fig:stab-diag-f}, where 
the resulting particle number pairs $(N_B,N_F)$ are smoothly connected with each other. The thick solid line of the 
Thomas-Fermi  ratio $\lambda_\mathrm{TF}=(\omega_z/\omega_r)^2$ lies above, but near the dot-dashed line from the 
Gross-Pitaevskii equation in the Thomas-Fermi approximation in Section \ref{Stab-T-F}. This is expected as both the 
thick solid line and the dot-dashed line are evaluated in the Thomas-Fermi limit with the same ratio of the radial 
and axial extension of the BEC cloud. Furthermore, they show the same behavior for very large boson numbers as both 
lines become vertical so that the fermion number remains constant. In the Florence experiment both lines lie very 
close to the crosses of the experiment where the thick solid line fits them better. This is not surprising as the 
s-wave scattering length $a_{BF}$ in Table \ref{tab:exp-parameter} is determined with a mean-field analysis in the 
Florence experiment \cite{Ferlaino}. The thin solid line lies far below the thick solid line and reflects a mixture 
in the quantum-mechanical limit with the ratio $\lambda_\mathrm{QM}=\omega_z/\omega_r$. This line, which is a good 
approximation for 
mixtures with vanishing intraspecies and interspecies interactions, is less suitable for the $^{87}$Rb--$^{40}$K mixture 
as the interaction energy is dominant according to Figure \ref{fig:energy}. Hence, the mixture can be well described 
in the Thomas-Fermi limit. As the thin solid line does not consider the proper ratio $\lambda$ of the oscillator 
lengths in 
the test function (\ref{test-func2}) and, thus, minimizes the free energy less optimally for a strong 
interacting mixture, 
this line allows much smaller particle numbers in a stable mixture than the thick solid line. In order to estimate the 
ratio $\lambda$, which yields the stability border with the largest possible numbers for bosons and fermions, we have 
evaluated two lines in the stability diagram for ratios in the neighborhood of $\lambda_\mathrm{TF}$, where one 
$\lambda$ is 40\% smaller  and the other one is 40\% larger. These lines for $\lambda=0.6\, \lambda_\mathrm{TF}$ and 
$\lambda=1.4\, \lambda_\mathrm{TF}$ 
are shown in Figure~\ref{fig:30-ab-bel}. Both lines lie throughout below, but very close to the original one with 
$\lambda= \lambda_\mathrm{TF}$. This indicates that the stability border in the Thomas-Fermi limit is stationary at 
$\lambda= \lambda_\mathrm{TF}$ where it possesses a maximum. Hence, the ratio $\lambda_\mathrm{TF}$ turns out to be the 
proper ratio 
for the $^{87}$Rb--$^{40}$K mixture allowing the largest possible numbers for bosons and fermions in a stable mixture.
\subsection{Expansion with respect to Interspecies Interaction}\label{expansion}
Another approach was suggested in Ref.~\cite{Chui}, where the local chemical potential (\ref{ferm-dens}) in the 
grand-canonical free energy (\ref{F9}) is expanded up to the third order in $g_{BF}$ in order to get rid off the 
fractional power in the last term of the free energy. This expansion leads to Gaussian integrals with respect to 
the test function (\ref{test-func2}):
\beq\label{F11}
\mathcal{F}[\Psi^*_B,\Psi_B] \hspace{-2mm}&=& \hspace{-2mm}\int d^3x\left[
\frac{\hbar^2}{2m_B}\,|\bs{\nabla}\Psi_B(\mathbf{x})|^2+
V_\mathrm{eff}({\mathbf{x}})\,|\Psi_B(\mathbf{x})|^2 + \frac{g_{\rm{eff}}}{2}\,|\Psi_B(\mathbf{x})|^4+ 
\frac{\kappa g_{BF}^3}{8\mu_F^{1/2}}\,|\Psi_B(\mathbf{x})|^6+\dots\right],
\eeq
where the terms with respect to the power of $|\Psi_B(\mathbf{x})|^2$ are summarized in the factors
\beq
V_\mathrm{eff}(\mathbf{x}) &=& \left[1-\frac{3}{2}\,\kappa\,\mu_F^{1/2}g_{BF}\right]\frac{1}{2}\,m_B\omega_{B,r}^2(r^2
+\lambda^2z^2),
\\
g_\mathrm{eff} &=& g_{BB}-\frac{3}{2}\,\kappa\,\mu_F^{1/2}g_{BF}^2.
\eeq
Within this approximation it is assumed that the radius of the condensate is much less than the radius of the Fermi 
gas cloud so that the remaining expressions $(\mu_F-V_F(\mathbf{x}))^{n/2}$ with $n=1,3,5$, arising from the expansion, 
can be expanded in powers of $V_F(\mathbf{x})/\mu_F$ as well. We consider therein only terms which depend on the 
parameter 
$\alpha$. The last term in the free energy (\ref{F11}) corresponds to the elastic three-particle collision induced by the 
interspecies interaction. This term with $g_{BF}<0$ is responsible for increasing both the bosonic and the fermionic 
particle density in the trap center in order to minimize the free energy. If the central condensate density 
$|\Psi_B(\mathbf{0})|^2$ becomes large enough due to large particle numbers, the positive first three terms in the free 
energy (\ref{F11}) cannot balance the negative last term in order to stabilize the mixture and to prevent it from 
collapsing. Performing the Gaussian integration in Eq.~(\ref{F11}) leads to an algebraic equation with respect to the 
unknown quantities $\alpha$ and $N_B$:
\beq\label{chui-algebr-eq}
\frac{\mathcal{F}(\alpha)}{N_B\hbar\omega_{B,r}} = \frac{2+\lambda}{4 \alpha^2}+\frac{b(2+\lambda)\alpha^2}{3}
+\frac{c_1N_B}{\alpha^3}
+\frac{c_2N_B^2}{\alpha^6}+\dots
\eeq
with the factors
\beq
b &=& \frac{3}{4}\left[1-\frac{3}{2}\,\kappa\,\mu_F^{1/2}g_{BF}\right],
\no\\
c_1 &=& \frac{1}{2}\left[g_{BB}-\frac{3}{2}\,\kappa\,\mu_F^{1/2}g_{BF}^2\right]\frac{\lambda^{1/2}}{(2\pi)^{3/2}\hbar
\omega_{B,r}L_{B,r}^3},
\no\\
c_2 &=& \frac{\kappa  \lambda g_{BF}^3}{3^{3/2} 8 \pi^3\hbar\omega_{B,r}\mu_F^{1/2}L_{B,r}^6}.
\eeq
The condition (\ref{stab-cond}) for the stability border provides two equations allowing to determine both unknown 
quantities for different values of $\mu_F$. The result is shown in Figures \ref{fig:stab-diag-h} and 
\ref{fig:stab-diag-f} 
by the dashed line for both limits. The thick dashed as well as the thin dashed line converges with increasing fermion 
number $N_F$ to the corresponding solid line of the numerical integration. But for low $N_F$ or, equivalently large 
$N_B$, these dashed lines stay below the solid lines, where the discrepancy increases with decreasing $N_F$. Moreover, 
for very small $N_F$ the dashed lines show the opposite behavior of the solid lines as they tend to zero, which seems 
to be unphysical. When the radius of the BEC clouds increases with decreasing $N_F$, the above mentioned expansion in 
powers of $V_F(\mathbf{x})/\mu_F$, which is done up to the zeroth and first order, fails as $V_F(\mathbf{x})$ and $\mu_F$ 
become comparable at the BEC cloud boundary. Thus, more orders of the expansion are needed to obtain more accurate 
results. Another reason is that the stability border in the stability diagram depends strongly on the interspecies 
s-wave scattering length $a_{BF}$ according to the scaling law for the critical numbers of condensate atoms for a 
fixed ratio between $N_B$ and $N_F$ \cite{Ferlaino,Molmer}:
\beq\label{N-a-depend}
N_{B\mathrm{crit}} \sim \frac{1}{a_{BF}^{12}}.
\eeq
Thus, such a strong 
sensitivity of the critical boson number $N_{B\mathrm{crit}}$ with respect to $a_{BF}$ and also to $g_{BF}$ 
due to Eq.~(\ref{g}) does not justify an expansion of the local chemical potential $\tilde \mu_F(\mathbf{x})$ with 
respect to the smallness parameter $g_{BF}$.
\begin{figure}[t]
\centerline{\includegraphics[scale=1.5]{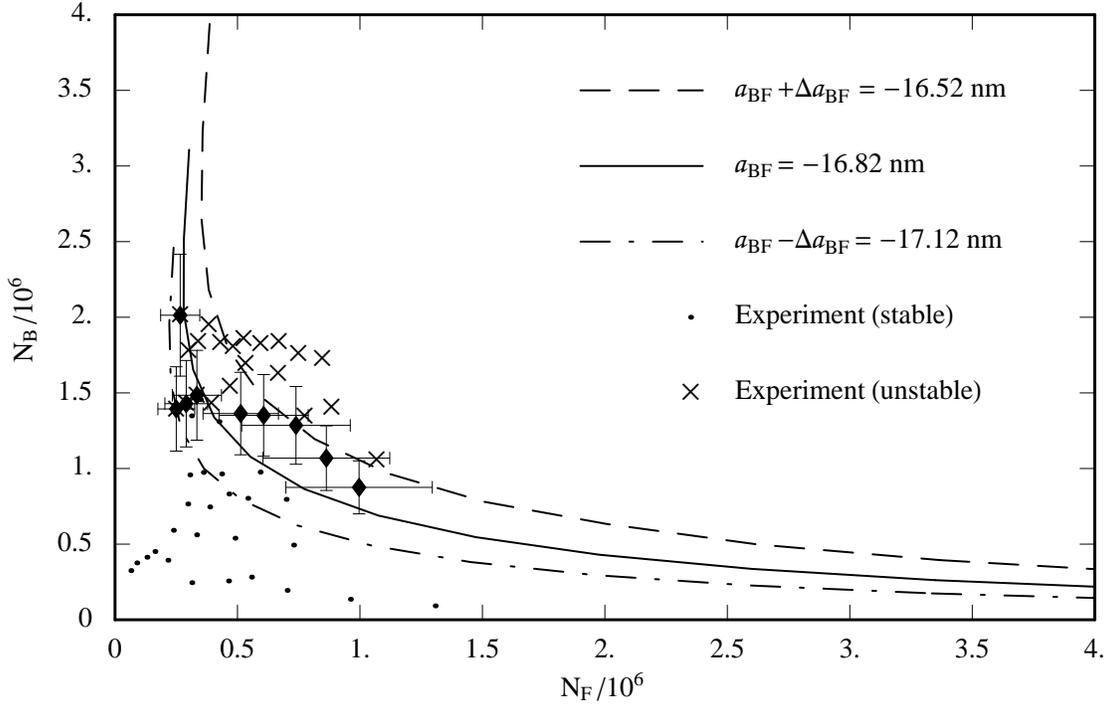}}
\caption{Stability border for a new value $a_{BF}=(-16.82\pm0.30)$ nm of the $^{87}$Rb--$^{40}$K mixture in the Hamburg 
experiment \cite{Sengstock}. All three lines are obtained by numerical integrations of the fermionic density 
(\ref{dens-F}) according to Subsection \ref{num-int} in the Thomas-Fermi limit with the ratio 
$\lambda_\mathrm{TF}=(\omega_z/\omega_r)^2$. Those stable (unstable) points, which are located above (below) the solid line, are equipped with error bars indicating a relative uncertainty of 20\% and 30\% for $N_B$ and $N_F$, respectively, according to Figure 3 of Ref.~\cite{Sengstock}.}
\label{fig:stab-diag-h1}
\end{figure}
\subsection{Adjustment of $a_{BF}$}\label{Stab-new-val}
The above described strong dependence of $N_{B\mathrm{crit}}$ on $a_{BF}$ allows us to extract a 
value for $a_{BF}$ with great accuracy within a mean-field analysis of the stability as performed in this section. 
Because of the scaling law (\ref{N-a-depend}) the relative uncertainty of $a_{BF}$ amounts to only a twelfth of the 
relative uncertainty of the critical boson number $N_{B\mathrm{crit}}$. As the thick solid line in Figure 
\ref{fig:stab-diag-f}, which is based on the value $a_{BF}=-20.9$ nm for the Florence experiment \cite{Ferlaino}, 
agrees quite well with the crosses of the experiments, we restrict ourselves here to fit 
$a_{BF}$ to the data of the Hamburg experiment. In order to estimate the uncertainty of the latter, 
we have plotted in Figure 
\ref{fig:stab-diag-h1} a dot-dashed (dashed) line comprising all crosses (dots) as a upper (lower) limit. The mean 
value for $a_{BF}$ is chosen central between both limits and is represented by the solid line in Figure 
\ref{fig:stab-diag-h1}, which roughly separates the dots and the crosses. This estimation method is 
justified since those stable (unstable) points, which are located above (below) the solid line, extend 
with their total error $\Delta N = \sqrt{\Delta N_B^2+\Delta N_F^2}$ over the solid line to the stable 
(unstable) sector. The new value $a_{BF}=(-16.82\pm 030)$ nm 
differs only 12\% from the old value $a_{BF}=-15.0$ nm, whereas the critical particle numbers between the dots and 
crosses are by a factor of around 3 smaller than those of the old thick solid line in Figure \ref{fig:stab-diag-h}.
Note that this new value for the interspecies s-wave scattering length leads also to a new value
of the minimal fermionic particle number $N_{F,{\rm min}}=1,95\,\cdot10^5$ below which
the mixture is stable in accordance with the experimental data of Figure~\ref{fig:stab-diag-h1}.   
Furthermore, we remark that the old value of $a_{BF}$
was determined by the Sengstock group with the help of Eq.~(\ref{chui-algebr-eq}). However, 
instead of evaluating the respective fermion number $N_F$ by integrating the fermionic density (\ref{dens-F}), they 
used the ad-hoc approximation \cite{Goldwinphd}
\beq
\mu_F = \mu_F^{(0)}-\frac{g_{BF}}{g_{BB}}\,\mu_B^{(0)},
\eeq
where the noninteracting chemical potentials $\mu_B^{(0)}$ and $\mu_F^{(0)}$ are related to the respective particle numbers by Eq.~(\ref{mueb}). The above relation between disturbed and undisturbed chemical potentials is not correct.
\section{Conclusion}\label{summary}
Applying the Thomas-Fermi approximation in order to solve the Gross-Pitaevskii equation (\ref{B-mot-eq-imag-ot}) 
for a stationary Bose-Fermi mixture in Section \ref{sol-GP} reduces it from a differential to an algebraic 
equation (\ref{algebr-GP-eq}). This allows to obtain an analytic expression for the bosonic and fermionic density 
profiles. The strong attraction between bosons and fermions gives rise to an increase of the particle densities 
within their overlap, accompanied by a shrinkage of the BEC and the Fermi gas cloud. In order to test the validity 
of the Thomas-Fermi approximation, we have plotted the kinetic energy of a boson together with its potential energy 
due to the trap, its intraspecies and its interspecies interaction energy, and the chemical potential as the total 
energy of a boson. This reveals the Thomas-Fermi approximation to be very good over a wide bulk range of the 
condensate cloud. The kinetic energy plays a significant role only in the outermost 10\% of the BEC cloud.

Furthermore, we have found that the particle densities for both species become complex at sufficiently large 
particle numbers of bosons and fermions. We interpret this as a loss of the stability against collapse. The 
imaginary part of the density can be regarded as the decay rate of the described species and starts emerging in
the trap center, where the densities have their maximum.

Beside the stability diagram arising from the complex solution of the Gross-Pitaevskii equation in the Thomas-Fermi 
approximation, we have evaluated the stability border within a variational method by extremizing the grand-canonical 
free energy (\ref{F9}) for a $^{87}$Rb--$^{40}$K mixture with a Gaussian test function. The resulting lines for the 
variational method with a ratio of the oscillator lengths according to the Thomas-Fermi approximation show the same 
behavior as the ones of the Gross-Pitaevskii equation. Both lines are located close in the $N_B$-$N_F$-plane and 
agree well. Finally, by comparing the calculated stability borders with the experimental values, we have found  
for the Florence experiment that the stability border of the variational method and of the Gross-Pitaevskii equation 
is in good agreement with the experimental results. For the Hamburg experiment, however, we have obtained a discrepancy 
between the calculated lines and the experimental results to which we have fitted the interspecies 
s-wave scattering length to $a_{BF}=(-16.82\pm0.30)$ nm. Despite this, there remains a discrepancy to the s-wave 
scattering length $a_{BF}=(-20.9\pm0.8)$ nm of the Florence experiment. As both experiments deal with a 
$^{87}$Rb--$^{40}$K 
mixture, the interspecies  s-wave scattering lengths should coincide.
A possible explanation could be that 
the mean-field theory in this paper is developed for a stationary Bose-Fermi mixture and does not include dynamical 
aspects of the mixture. The dynamical behavior becomes important for a rapid cooling of the mixture which leads to a 
fast increase of the condensate density due to the occupation of all bosons into the ground state in a short time. We 
conclude with the remark that 
such dynamical effects within 
the Bose-Fermi mixture could be further investigated with the help of the coupled equations of 
motion (\ref{B-mot-eq-imag}) and (\ref{F-mot-eq-imag}).
\section*{Acknowledgement}
We thank Kai Bongs, Konstantin Glaum, and Aristeu Lima for critical comments as well as
the DFG Priority Program SPP 1116 {\it Interactions in Ultra-Cold Atomic and 
Molecular Gases} for financial support.    

\end{document}